\documentclass[12pt,fleqn]{article}
\usepackage{epsfig}
\usepackage{amsfonts}
\usepackage{amsmath}
\usepackage{amssymb,latexsym}
\usepackage{amstext}
\usepackage{amsthm}
\usepackage{graphics,keyval,graphicx}
\numberwithin{equation}{section}
\begin{document}

\title{\Large {\bf{ Euler potentials for the MHD Kamchatnov-Hopf soliton
solution}}}

\author{Vladimir S. Semenov$^{\mbox{1}}$,
 Daniil B. Korovinski$^{\mbox{1}}$,
 and Helfried K.Biernat$^{\mbox{2}}$}
 \date{}
\maketitle

\begin{flushleft}
\emph{$^{\mbox{1}}$Institute of Physics, State University, St.Petersburg, 198504 Russia} \\
\emph{$^{\mbox{2}}$Space Research Institute, Austrian Academy of
Sciences, Schmiedlstrasse 6, A--8042 Graz, Austria}
\end{flushleft}


\begin{abstract}                

In the MHD description of plasma phenomena the concept of magnetic
helicity turns out to be very useful. We present here an example
of introducing Euler potentials into a topological MHD soliton
which has non-trivial helicity. The MHD soliton  solution
(Kamchatnov, 1982) is based on the Hopf invariant of the mapping
of a 3D sphere into a 2D sphere; it can have arbitrary helicity
depending on control parameters. It is shown how to define Euler
potentials globally. The singular curve of the Euler potential
plays the key role in computing helicity. With the introduction of
Euler potentials, the helicity can be calculated as an integral
over the surface bounded by this singular curve. A special
programme for visualization is worked out. Helicity coordinates
are introduced which  can be useful  for numerical simulations
where helicity control is needed.

\end{abstract}

\section{Introduction}

Magnetic helicity is a topological characteristic of magnetic
field structures which includes the twisting and the kinking of a
flux tube as well as the linkage between different flux tubes
(Moffatt, 1978, Biskamp, 1993). Among its numerous applications
are dynamo theory (Moffatt, 1978), investigation of magnetic
reconnection (Wiegelmann and B\"uchner, 2001), theory of
relaxation (Taylor, 2000), and even the collimation mechanism of
astronomical jets (Yoshizawa et al., 2000). Magnetic helicity is
defined as a volume integral

\begin{equation}\label{helicity}
K=\int_\Omega(\mathbf{A}\cdot\mathbf{H})dV
\end{equation}

where $\mathbf{B}$ is the magnetic field and $\mathbf{A}$ is the
vector potential

\begin{equation}\label{A}
\mathbf{B}=\mathbf{\nabla} \times \mathbf{A}.
\end{equation}

Helicity (\ref{helicity}) is gauge invariant, because under the
transformation $\mathbf{A'}\rightarrow \mathbf{A}+\nabla \phi$, it
then is changed by

\begin{equation}\label{calibr}
 \delta K=\int_\Omega(\nabla\phi\cdot{\mathbf{B})}\,d^3x=
 \oint_{\partial\Omega}\phi({\mathbf{B}\cdot d\mathbf{S}})=0,
 \end{equation}

if ${B}_n|_{\partial\Omega}=0$, where ${\mathbf{n}}$ is the vector
normal to the boundary $\partial\Omega$. For
${B}_n|_{\partial\Omega}\ne 0$ the surface integral does not
vanish and the helicity becomes gauge dependent. Generally
speaking, there is the possibility to define the helicity for
difference between the original field and the vacuum field
(Schindler et al., 1988; Biskamp, 1993; Priest and Forbes, 2000)
which helps to give the helicity a physical meaning for more
realistic conditions. Nevertheless, we will restrict our
consideration to the classical case ${B}_n|_{\partial\Omega}=0$,
leaving a more general definition of the magnetic helicity for
future studies.
When the Euler potentials $\alpha, \, \beta$ are
used,

\begin{equation}\label{euler}
\mathbf{B}=\nabla\alpha\times\nabla\beta,
\end{equation}

there is the following problem related to helicity. It can be
easily verified that

\begin{equation}\label{Abad}
\mathbf{A}=-\beta\nabla\alpha
\end{equation}

(or $\mathbf{A}=\alpha\nabla\beta$) is the vector potential
(\ref{A}) for the magnetic field (\ref{euler}). Then helicity
vanishes at the level of the scalar product
$({\mathbf{A}\cdot\mathbf{B}})=0$. It is known (see, for example,
Biskamp, 1993) that the vector potential can be presented in the
following form (Clebsch representation)

\begin{equation}\label{Clebsch}
\mathbf{A}=-\beta\nabla\alpha+\nabla\psi,
\end{equation}

where the function $\psi$ (contrary to $\phi$ in (\ref{calibr}))
must be multi-valued. This implies that the function $\psi$ has a
surface $S_j$ inside the volume $\Omega$ where it has a jump, then
the contribution from the jump surface $S_j$ is added to the
integral over $\partial\Omega$ in  equation (\ref{calibr}) which
results in the nonzero helicity. The solution to the questions how
to introduce Euler potentials globally for the magnetic field with
non-trivial helicity, how to find the function $\psi$, and why it
has to be multi-valued, are not clear so far. For example, it is
stated (Biskamp, 1993) that Euler potentials can not be introduced
globally for a magnetic field with nonzero helicity unless the
system is multiply connected. In (Sagdeev et al., 1986) it is
pointed out that magnetic field lines determined by the Lagrangian
invariants do not admit any linkage, i.e.,the helicity has to
vanish. The representation (\ref{Abad}) is used sometimes (Priest
and Forbes, 1999; Wong, 2000) quite generally, but it is not
mentioned that helicity has to vanish in this case, hence the
structure of the magnetic field has to be relatively simple.

 The aim of this paper is to
show how one can practically introduce Euler potentials
(\ref{euler}) as well as the Clebsch representation
(\ref{Clebsch}) for a magnetic field with nonzero helicity. There
is a solution of the MHD equations (Kamchatnov, 1982, Sagdeev et
al., 1986) based on the Hopf invariant of the mapping of a 3D
sphere $S^3$ into a 2D sphere $S^2$ (see, for example, Dubrovin et
al., 1979). In this solution the magnetic flux tubes can link each
other as many times as one wants. The MHD soliton has a known
helicity following from topology, hence in each step of the
calculation, there is opportunity to control the situation.
Besides, this solution is relatively simple, and all the results
can be obtained analytically. We will not use topological methods,
because all our results can be obtained straightforwardly if some
topological information has been taken into account from the very
beginning.

This paper is organized as follows. In Sections 2 and 3 we recall
the details of the MHD Kamchatnov-Hopf solution. Euler potentials
are introduced in Section 4. A visualization of the magnetic field
structure is presented  in Section 5. Helicity coordinates are
introduced in Section 6,  and Section 7 is devoted to the summary
and discussion.

\section{MHD Kamchatnov-Hopf soliton}

First of all we will recall (Chandrasekhar, 1961; Kamchatnov,
1982) that any solinoidal vector field,
$\mbox{div}{\mathbf{B}}=0$, gives rise to a solution of the
steady-state MHD equations

\begin{align}
& \rho {(\mathbf{v}\cdot\nabla)\mathbf{v}=-\nabla} P
+\frac{1}{4\pi}(\mathbf{B}\cdot\nabla)\mathbf{B},  \label{motion}
\\
&(\mathbf{v}\cdot\nabla)\mathbf{B}=(\mathbf{B}\cdot\nabla)\mathbf{v},
\label{vb}
\\
&\mbox{div}\mathbf{v}=0 , \label{dv}
 \\
&\mbox{div}\mathbf{B}=0, \label{db}
\end{align}

in an incompressible plasma where the density $\rho=const$. Here
$P$ is the total (gas + magnetic) pressure, $\mathbf{v}$ is the
plasma velocity. If we choose ${\mathbf{v}}={\mathbf{B}}/{\sqrt{4
\pi \rho}}$, and $P=const$, then equations (\ref{motion} -
\ref{db}) are satisfied automatically. In this solution the
magnetic tension is balanced by the centrifugal force.

The idea of the Kamchatnov-Hopf soliton solution is to obtain a
solenoidal vector field with known linkage using topological
methods. A 3D sphere $S^3$ is defined in $R^4$ as a set of points
$(q_1,q_2,q_3,q_4)$ such that $q_1^2+q_2^2+q_3^2+q_4^2=1$. Let us
introduce two complex numbers $Z_{1}=q_{1}+iq_{2},
Z_{2}=q_{3}+iq_{4}$, then $S^3$ can be described also as
$|Z_{1}|^2+|Z_{2}|^2=1$. A curve (a circle) on $S^3$ can be
presented as

\begin{equation}\label{curve}
{\mathbf{l}}(t)=(Z_1 e^{i\omega_1t},Z_2 e^{i\omega_2t}),
\end{equation}

where $t$ is a parameter along the curve.  It can be shown
(Dubrovin et al., 1979) that two curves corresponding different
initial points $Z_{1},Z_{2}$ with integer numbers $\omega_1,
\omega_2$  link each other $\omega_1 \omega_2$ times.

A tangential field $\mathbf{Y}$ on $S^3$ generated by the curve
(\ref{curve}) is

\begin{equation}\label{fieldY}
{\mathbf{Y}}(\omega_1,\omega_2)=\frac{d{\mathbf{l}}(t)}{dt}=
(-\omega_1q_2,\omega_1q_1,-\omega_2q_4,\omega_2q_3),
\end{equation}

which also has the linkage $\omega_1 \omega_2$. Now we can map the
curve (\ref{curve}) into $R^3$ using the stereographic projection

\begin{equation}\label{xfq}
x_i=\frac{q_i}{1+q_4},\quad i=1,2,3,
\end{equation}

\begin{equation}\label{qfx}
q_4=\frac{1-x^2}{1+x^2},\quad q_i=\frac{2x_i}{1+x^2},\quad
i=1,2,3.
\end{equation}

To obtain the vector field (\ref{fieldY}) in $R^3$ we can just
differentiate equation (\ref{xfq}) with respect to parameter $t$

\begin{equation}\label{fieldJ}
{\mathbf{J}}=\biggl[-(\omega_{2}x_{1}x_{3}
+\omega_{1}x_{2}),\quad(\omega_{1}x_{1}
-\omega_{2}x_{2}x_{3}),\quad\frac{1}{2}\omega_{2}
(x_{1}^{2}+x_{2}^{2}-x_{3}^{2}-1)\biggl].
\end{equation}

Stereographic projection conserves the topological invariant that
is the linkage $\omega_1 \omega_2$.

As a matter of fact, $\mbox{div}{\mathbf{J}}\ne 0$, but it can be
easily verified that the field
${\mathbf{B}}=4*{\mathbf{J}}/{(1+x^2)^3}$ is solenoidal, where
$x^2=x_1^2+x_2^2+x_3^2$. The factor $1/(1+x^2)^3\ne 0$ everywhere
in $R^3$, therefore the field obtained,

\begin{equation}\label{B}
{\mathbf{B}}=\frac{2}{(1+x^2)^3}\{-2(\omega_{2}x_{1}x_{3}
+\omega_{1}x_{2}),\quad 2(\omega_{1}x_{1}
-\omega_{2}x_{2}x_{3}),\quad \omega_{2}
(x_{1}^{2}+x_{2}^{2}-x_{3}^{2}-1)\},
\end{equation}

has the same topological property as the field (\ref{fieldY}) on
$S^3$. The factor $4$ was introduced for the calculations
convenience.

The field (\ref{B}) is the basis for the topological soliton. As
was pointed out previously, if we introduce the plasma velocity
${\mathbf{v}}={\mathbf{B}}/\sqrt{4 \pi \rho}$, and the pressure
$P=const$, then MHD equations (\ref{motion} - \ref{db}) are
satisfied automatically. We will refer to this solution as the MHD
Kamchatnov-Hopf soliton.

\section{Magnetic field lines.}

Let us now derive the equation of the magnetic field lines in
$R^3$. To this end we can solve differential equations $\frac{d
{\mathbf{r}}}{d \lambda}=\mathbf{B}$ using (\ref{B}), but it is
much more easy just to map the known integral curves (\ref{curve})
from $S^3$ to $R^3$ with the help of stereografic projection
(Kamchatnov, 1982; Sagdeev et al., 1986)















\begin{align}\label{x(t)}
x_1(t)&=\frac{2(x_{10}\cos(\omega_1t)-x_{20}\sin(\omega_1t))}
{1+x_0^2+(1-x_0^2)\cos(\omega_2t)+2x_{30}\sin(\omega_2t)},
\nonumber
\\
x_2(t)&=\frac{2(x_{20}\cos(\omega_1t)+x_{10}\sin(\omega_1t))}
{1+x_0^2+(1-x_0^2)\cos(\omega_2t)+2x_{30}\sin(\omega_2t)},
\\
x_3(t)&=\frac{2x_{30}\cos(\omega_2t)-(1-x_0^2)\sin(\omega_2t)}
{1+x_0^2+(1-x_0^2)\cos(\omega_2t)+2x_{30}\sin(\omega_2t)}.
\nonumber
\end{align}

Using trigonometric identities, it is possible to reduce equations
(\ref{x(t)}) to the following form

\begin{align}\label{x}
x_1&=\frac{\cos\Theta_1}{a+b\cos\Theta_2}, \nonumber
\\
x_2&=\frac{\sin\Theta_1}{a+b\cos\Theta_2},
\\
x_3&=\frac{b\sin\Theta_2}{a+b\cos\Theta_2}, \nonumber
\end{align}

where

\begin{align}\label{xx}
&\Theta_1=\omega_1t+\alpha_1,\quad
\Theta_2=-\omega_2t+\alpha_2,\quad
a=\frac{1+x_0^2}{2\sqrt{x_{10}^2+x_{20}^2}},\nonumber
\\
&b^2=a^2-1=\frac{4x_{30}^2+(1-x_0^2)}{4(x_{10}^2+x_{20}^2)},\quad
\cos\alpha_1=\frac{x_{10}}{\sqrt{x_{10}^2+x_{20}^2}},\nonumber
\\
&\sin\alpha_1=\frac{x_{20}}{\sqrt{x_{10}^2+x_{20}^2}}, \quad
\cos\alpha_2=\frac{1-x_0^2}{\sqrt{4x_{30}^2+(1-x_0^2)^2}},\quad
\sin\alpha_2=\frac{2x_{30}}{\sqrt{4x_{30}^2+(1-x_0^2)^2}}.
\end{align}

It turns out that the magnetic field lines lie on the surface of
the torus

\begin{align}\label{torus}
&x_1=(a+b\cos\Theta_2)\cos\Theta_1,\nonumber
\\
&  x_2=(a+b\cos\Theta_2)\sin\Theta_1,
\\
&x_3=b\sin\Theta_2,\nonumber
\end{align}

which is produced by the rotation of the circle
$x_3^2+(x_1-a)^2=a^2-1$ around the $x_3$ axis. The central torus
degenerates into a circle (Sagdeev et. al, 1986),

\begin{equation}\label{circle}
x_3=0,\quad x_1^2+x_2^2=1,
\end{equation}

which will play an important role hereafter.

\section{Euler potentials}

It is convenient to choose as Euler potentials the following
constants of integration (first integrals) of (\ref{xx})

\begin{align}\label{alpha}
\alpha =\alpha_1\omega_2+\alpha_2\omega_1, \quad \beta
=\frac{1}{(2a)^2},
\end{align}

or, in Cartesian coordinates,

\begin{align}\label{beta}
&\alpha
=\omega_2\arctan\frac{x_2}{x_1}+\omega_1\arctan\frac{2x_3}{1-x^2},
\\
& \beta =\frac{x_1^2+x_2^2}{(1+x^2)^2}.\nonumber
\end{align}

Then we can find the gradients of these functions

\begin{equation}\label{grada}
\nabla\alpha=\biggl[\frac{4\omega_{1}x_{1}x_{3}}{(x^{2}-1)^{2}+x_{3}^{2}}-\frac{\omega_{2}x_{2}}{x_{1}^{2}+x_{2}^{2}},\quad
\frac{4\omega_{1}x_{2}x_{3}}{(x^{2}-1)^{2}+x_{3}^{2}}+\frac{\omega_{2}x_{1}}{x_{1}^{2}+x_{2}^{2}},\quad
-\frac{2\omega_{1}(x_{1}^{2}+x_{2}^{2}-x_{3}^{2}-1)}{(x^{2}-1)^{2}+x_{3}^{2}}\biggl],
\end{equation}

\begin{equation}\label{gradb}
\nabla\beta=-\frac{2}{(1+x^2)^3}\;[x_1(x_1^2+x_2^2-x_3^2-1),\quad
x_2(x_1^2+x_2^2-x_3^2-1),\quad 2x_3(x_1^2+x_2^2)],
\end{equation}

and verify that equation (\ref{euler}) is satisfied, i.e., the
$\alpha,\beta$ are indeed  Euler potentials. The potential
$\alpha$ is a naked angle (i.e., an angle being not hidden  under
any trigonometric functions), which can have a nonzero
contribution after integration of its gradient along a closed
contour. Therefore, it is not surprising that first of all,
$\alpha$ is a multi-valued function, and secondly, $\nabla\alpha$
has a singularity on the circle (\ref{circle}).

The next step is to obtain the vector potential
$\mathbf{{A}}=-\beta\nabla\alpha$

\begin{align}\label{bgra}
&{\mathbf{A}}=\biggl\{\frac{-4\omega_{1}x_{1}x_{3}(x_{1}^{2}+x_{2}^{2})}{R}
+\frac{\omega_{2}x_{2}}{(x^{2}+1)^{2}}, \nonumber
\\
&\frac{-4\omega_{1}x_{2}x_{3}(x_{1}^{2}+x_{2}^{2})}{R}-\frac{\omega_{2}x_{1}}{(x^{2}+1)^{2}},
\\
&\frac{2\omega_{1}(x_{1}^{2}+x_{2}^{2})(x_{1}^{2}+x_{2}^{2}-x_{3}^{2}-1)}{R}\biggl\},\nonumber
\end{align}

where $R=(x^{2}+1)^{2}((x^{2}-1)^{2}+4x_{3}^{2})$. Remember that
the formal representation (\ref{Abad}) leads to zero helicity, but
we know that $K\ne0$ by the topological construction, hence, the
potential (\ref{bgra}) should have some principal disadvantage. If
vector potential $\mathbf{A}$ is defined by the differential
equation (\ref{A}), then we have to conclude that $\mathbf{A}$ is
indeed the vector potential of the magnetic field $\mathbf{B}$
because the equation (\ref{A}) is satisfied. But besides  the
differential equation there is also an integral equation

\begin{equation}\label{Aint}
\oint_L({\mathbf{A}\cdot d\mathbf{l}})=\int_S({\mathbf{B}\cdot
d\mathbf{S}})=F_B,
\end{equation}

where $F_B$ is the magnetic flux through the surface $S$ bounded
by the contour $L$. Differential and integral equations sometimes
are not identical, and in our situation this is exactly the case.
If we choose any contour $L$ which does not cross the disc bounded
by the singular circle (\ref{circle}), then the circulation of
$\mathbf{A}$ along $L$ gives exactly the magnetic flux $F_B$.
However, if the contour encounters the disc bounded by singular
circle (\ref{circle}), then the circulation gets an additional
contribution

\begin{equation}\label{Aint1}
\oint_L({\mathbf{A}\cdot d\mathbf{l}})= F_B+\frac{\pi\omega_1}{2},
\end{equation}

therefore, the integral equation (\ref{Aint}) is not satisfied.
The formal reason for the multi-valued character of the
circulation (\ref{Aint1}) lies in the singular behaviour of the
latter at the circle (\ref{circle}), or due to the fact that the
function $\alpha$ (\ref{alpha}) is a naked angle. Hence, we have
to proceed with the Clebsch representation (\ref{Clebsch}), and to
find a function $\psi$ to compensate the singularity in the
potential (\ref{Abad}). It is clear that the function $\psi$ has
also to be  a naked angle like the function $\alpha$, and its
gradient should have a singularity at the circle (\ref{circle})

\begin{equation}\label{psi}
\psi=\frac{1}{4}\omega_{1}\arctan(\frac{x^{2}-1}{2x_{3}})=
\frac{1}{4}\omega_1(-\omega_2t+\alpha_2+\frac{\pi}{2}).
\end{equation}

Then, the Clebsch potential (\ref{Clebsch}) turns out to be

\begin{equation}\label{Agood}
{\mathbf{A}}=\biggl\{\frac{\omega_{2}x_{2}+\omega_{1}x_{1}x_{3}}{(1+x^{2})^{2}},\quad
\frac{\omega_{1}x_{2}x_{3}-\omega_{2}x_{1}}{(1+x^{2})^{2}},\quad
\frac{\omega_{1}(1+x_{3}^{2}-x_{1}^{2}-x_{2}^{2})}{2(1+x^{2})^{2}}\biggl\}.
\end{equation}

It has no singularity in the whole space like the magnetic field
(\ref{B}), and both the differential (\ref{A}) and the integral
(\ref{Aint}) equations are now satisfied.

It is interesting to note that the Clebsch representation
(\ref{Clebsch}) formally looks similar to the gauge condition
$\mathbf{A'}\rightarrow \mathbf{A}+\nabla \phi$. Nevertheless
there is an essential difference. The function $\phi$ has to be a
single-valued one for the gauge transformation at least for the
simple connected region $\Omega$, hence the integral of its
gradient along any closed contour has to vanish. Contrary, the
function $\psi$ in the Clebsch representation (\ref{Clebsch}) has
to be a multi-valued one, and the integral of its gradient along
some closed contour can have nonzero contribution. Generally
speaking, the question whether the gauge function is a multi or
single valued one is not really important for many applications in
electrodynamics. But for such a delicate characteristics      of
the field as the magnetic helicity, the solution of this question
plays the key role. It is the multi-valued function $\psi$ which
does the nonzero helicity.

Using the vector potential (\ref{Agood}) and the magnetic field
(\ref{B}), we can calculate the helicity  as the volume integral
(\ref{helicity})

\begin{equation}\label{K}
K=-\frac{\pi^{2}\omega_{1}\omega_{2}}{4}.
\end{equation}

The negative sign in (\ref{K}) is connected with the parameter $t$
in the initial curve (\ref{curve}) at $S^3$, so that $e^{i\omega
t}$ gives a minus, whereas $e^{-i\omega t}$ gives a plus.

The Clebsch representation (\ref{Clebsch}) leads to another way to
compute the helicity

\begin{equation}\label{kpsi}
K=\{\psi\}\int_S(\mathbf{B}\cdot \,dS),
\end{equation}

which, of course gives the same result (\ref{K}). Here $S$ is the
singular circle (\ref{circle}), and $\{\psi\}$ is the jump of the
function $\psi$ on the latter. As one can see, helicity can be
calculated from the surface integral (\ref{kpsi}) rather than from
the  volume integral (\ref{helicity}), which is simpler to do. It
is also interesting that the helicity is equal to the magnetic
flux through the singular circle times the jump of the $\psi$
function.

\section{Visualization }

It is worthwhile to present pictures of the magnetic field
structure of the MHD Kamchatnov-Hopf soliton solution as
mathematical examples for illustration.

We start with the simplest case, $\omega_1=\omega_2=1$ . The flux
tube looks like a torus twisted by the angle $360^o$. To see this
more clearly, the tube presented is chosen to have a rectangular
cross section (Figure 1) so that one can easily  follow the
screwed color boundaries.

The surface Euler potential  $\beta=const$ is just a usual torus
(Figure 2), it stays more or less the same for all $\omega_1,
\omega_2$. The magnetic field lines are swept around this torus.

The surface $\alpha=const$ is more complicated (Figure 3). It is
similar to a ribbon twisted  by $360^o$. Such a surface can not be
continued to the closed one in $R^3$ without self crossing which
is because $\nabla\alpha$ has a singularity at the circle
(\ref{circle}).

There is a simple way to imagine the magnetic field structure. Let
us take a paper ribbon, twist it by the angle $360^o$ (note that
twisting by $180^o$ gives a Moebius sheet), glue the edges of the
ribbon together, and then cut it along the central line with
scissors. As a result, we get two ribbons linked to each other. If
we continue this procedure and cut the two ribbons obtained along
their central axis, and so on, we can observe that each ribbon
links any other one exactly one time (Figure 4). This behaviour is
reflected in the topological invariant helicity $K$ (\ref{K}).

It is difficult to imagine that the intersection of two surfaces
$\alpha=const$ and $\beta=const$ for different constants can give
linked lines, nevertheless, it is so.

To complete the case $\omega_1=\omega_2=1$, we present also the
surface $\psi=const$ (Figure 5) which has a spiral structure
converging to the singular circle (\ref{circle}).

After these relatively simple pictures we can proceed to the
general case. First we recall that two numbers are relatively
prime, if and only if the greatest common divisor of the numbers,
is one. For integers $\omega_1, \omega_2$ such that $\omega_1=n,
\omega_2=m$ are relative prime, the magnetic field lines of the
MHD Kamchatnov-Hopf soliton are linked into $(n, m)$ knots which
are topologically nonequivalent for different $(n, m)$. They form
the known family of toric nodes (Crowell and Fox, 1963).

The case $\omega_1=2, \omega_2=1$  is depicted in Figure 6 (single
flux tube), Figure 7 (surface $\alpha=const$), and Figure 8 (knot
(2,1) ). The more complicated case $\omega_1=2, \omega_2=3$ is
presented in Figure 9 (single flux tube), Figure 10 (surface
$\alpha=const$), Figure 11 (central fragment of the surface
$\alpha=const$), and Figure 12 (knot (2,3) ).

It is interesting that the surface of the Euler potential
$\alpha=const$ for the latter case (Figures 10, 11) is similar to
a propeller, and this circumstance seems not to be a pure
coincidence. The propeller has to create curls of air for
producing a moving force, and at least some surfaces
$\alpha=const$ (Figures 3, 7, 10) might be used for this aim just
from  topological reasons. Of course there is the question about
the efficiency of such airscrews or waterscrews, but this is not a
subject of this paper.

\section{Helicity coordinates}

A  magnetic field line is defined by two Euler potentials $\alpha,
\beta$, and a point on this line is controlled by the parameter
$t$. We can use another parameter $\psi$ along the magnetic field
line instead of $t$. Then $\alpha,\beta, \psi$, i.e., all
functions taking part in the Clebsch representation of the vector
potential (\ref{Clebsch}), can be used as new curvilinear
coordinates which have some useful property.

We already have an expression for the Clebsch coordinates via
Cartesian coordinates (\ref{alpha}, \ref{beta}, \ref{psi}). It is
possible to simplify these equations noting that without loss of
generality, we can assume $\alpha_1=0$ in (\ref{alpha}) and then
obtain

\begin{align}\label{hcoor1}
&\alpha =\omega_1\arctan\frac{2x_3}{1-x^2},\nonumber
\\
&\beta =\frac{x_1^2+x_2^2}{(1+x^2)^2},
\\
&\psi=\frac{1}{4}\omega_{1}\arctan(\frac{x^{2}-1}{2x_{3}}).\nonumber
\end{align}

Now we can also find the mapping $(x_1,x_2,x_3)\rightarrow
(\alpha, \beta, \psi)$

\begin{align}\label{hcoor2}
&x_1=\frac{2\sqrt{\beta}\cos(\frac{\alpha}{\omega_2}+
\frac{\pi\omega_1}{2\omega_2}-\frac{4\psi}{\omega_2})}
{\sqrt{1-4\beta}\cos(\frac{4\psi}{\omega_1}-\frac{\pi}{2})+1},\nonumber
\\
&x_2=\frac{2\sqrt{\beta}\sin(\frac{\alpha}{\omega_2}+
\frac{\pi\omega_1}{2\omega_2}-\frac{4\psi}{\omega_2})}
{\sqrt{1-4\beta}\cos(\frac{4\psi}{\omega_1}-\frac{\pi}{2})+1},
\\
&x_3=\frac{\sqrt{1-4\beta}\sin(\frac{4\psi}{\omega_1}
-\frac{\pi}{2})}{\sqrt{1-4\beta}\cos(\frac{4\psi}{\omega_1}
-\frac{\pi}{2})+1}.\nonumber
\end{align}

After some algebra one can find the Jacobian of this
transformation

\begin{equation}\label{J1}
J=\frac{D(x_1,x_2,x_3)}{D(\alpha,\beta,\psi)}= \frac{A}{B},
\end{equation}

where

\begin{align}\label{J2}
&A=8(\sqrt{(1-4\beta)}+4\beta\gamma^2\sqrt{(1-4\beta)}
+2(1-4\beta)^{\frac{3}{2}}\gamma^2 -8\beta\gamma
+2\gamma-\sqrt{(1-4\beta)\gamma^2)}, \nonumber
\\
&B=\omega_2\omega_1(-\gamma^5+
12\beta\gamma^5-48\beta^2\gamma^5+64\beta^3\gamma^5-
5(1-4\beta)^{\frac{5}{2}}\gamma^4-10\gamma^3+ 80\beta\gamma^3-
160\beta^2\gamma^3 \nonumber
\\
&-10(1-4\beta)^{\frac{3}{2}}\gamma^2- 5\gamma+
20\beta\gamma-\sqrt{(1-4\beta)}),
\\
&\gamma=\cos(-4\frac{\psi}{\omega_1}+\frac{\pi}{2}).\nonumber
\end{align}

This equation is a bit complicated, nevertheless it is possible to
verify that $J\ne0$ in the whole space, hence the coordinates
$(\alpha,\beta,\psi)$ can be introduced in $R^3$.

Let us compute the magnetic helicity using these new coordinates

\begin{align}\label{hell}
&K=\int_{R^3}({\mathbf{A}\cdot\mathbf{B}})dx_1\wedge dx_2\wedge
dx_3= \int_{R^3} (\nabla\psi \cdot [\nabla\alpha \times
\nabla\beta])dx_1\wedge dx_2\wedge dx_3=\nonumber
\\
&\int_{R^3} \frac{D(\alpha,\beta,\psi)}{D(x_1,x_2,x_3)}dx_1\wedge
dx_2\wedge dx_3=\int_{\Omega'} d\alpha d\beta d\psi,
\end{align}

where it is supposed that the whole space is mapped into the
region $\Omega'$, $R^3\rightarrow\Omega'$ under the transformation
$(x_1,x_2,x_3)\rightarrow (\alpha, \beta, \psi)$. Therefore, it
turns out that in the new variables, magnetic helicity is equal to
the volume of the configuration space $(\alpha,\beta,\psi)$. It is
easy to verify that the new formula (\ref{hell}) gives the same
result (\ref{K}) for the helicity if we take into account that the
coordinates $(\alpha,\beta,\psi)$ are varied within the following
limits

\begin{align}\label{prl}
\alpha\in(-\pi,\pi];\quad \beta\in
[0,\frac{1}{4});\quad\psi\in(-\frac{1}{4}\omega_1\omega_2\pi,
\frac{1}{4}\omega_1\omega_2\pi],
\end{align}

where brackets ( or [ are used to show that the element close to
the bracket is excluded or included in the list of elements,
respectively.

 One can see that the space $R^3$ is mapped onto the
parallelepiped (\ref{prl}) in which the straight lines
$(\alpha=const,\beta=const)$ represent  the magnetic field lines.
It is surprising that all the complicated magnetic structure is
converted into a very simple geometrical object, that is the
parallelepiped (\ref{prl}). In fact, the situation is not that
simple. To make field lines which are closed, we have to glue the
end points. The points on the left boundary $\alpha=-\pi$ have to
be considered identical with those on the right boundary
$\alpha=\pi$, and the points of the bottom boundary
$\psi=-\frac{1}{4}\omega_1\omega_2\pi$ are identical with those on
the upper boundary $\psi=\frac{1}{4}\omega_1\omega_2\pi$ after the
rotation of the latter by the  angle $2\omega_1\omega_2\pi$.

We note that helicity coordinates can be  particularly important
for numerical simulations where helicity control is required.

\section{Discussion and summary}

It was shown that  Euler potentials can be introduced globally for
a magnetic field with nonzero helicity even for the
simply-connected region (the space $R^3$ in our case), contrary to
the remark of Biskamp (1993). Therefore most of the coordinate
systems (Pudovkin and Semenov, 1985; Pustovitov, 1999) based on
Euler potentials (\ref{euler}), such as the helicity system
(\ref{hcoor2}) can still  be applied also to magnetic structures
with nonzero helicity $K\ne0$. On the other hand, one has to be
particularly  careful with the vector potential. Remember that the
simple representation (\ref{Abad}) is not appropriate for the
magnetic field with $K\ne0$, instead the Clebsch representation
(\ref{Clebsch}) has to be used.

As we saw,  the function $\psi$ plays a key role in calculating
the magnetic helicity. The multi-valued character of this function
is connected with the singular behaviour of the gradient of at
least one Euler potential ($\alpha$ in our case). In its turn, the
singularity of the Euler potential is the consequence of the fact
that the $\alpha=const$ surface is highly twisted for the case
$K\ne0$ and cannot be continued to a closed surface in $R^3$.

The helicity turns out to be equal to the magnetic flux through
the singular circle times the jump of the function $\psi$, hence,
the calculation of $K$ can be reduced  to a surface integral.It
seems that the simple formula (\ref{kpsi}) can be extended to the
general case as

\begin{equation}
K=\sum_j \{\psi_j\} F_{Bj},\label{kpsi1}
\end{equation}

where $F_{Bj}$ is the magnetic flux through the surface bounded by
$j$-singular curve of the Euler potential and $\{\psi_j\}$ is the
jump of the function $\psi$ at this surface.

If by chance it is known that all singular lines of the Euler
potentials lie on  a surface $S$  (the surface of the Sun, for
example) then the magnetic helicity can be found using only data
of the normal component of the magnetic field $B_n$ on $S$ from
the equations (\ref{kpsi} or \ref{kpsi1}) which is an important
problem for solar physics. But if a singular line is inside the
Sun, it is not possible to find the helicity using surface data.
The maximum of what can be done is to estimate the helicity if one
could somehow control  the magnetic flux closed under the Sun's
surface.

 The Kamchatnov-Hopf solution seems to be the simplest one which can describe the
magnetic field with such a non-trivial helicity. Therefore it may
play the same role for the investigation of different helicity
problems as the Harris (1963) layer in plasma physics or the
Petschek(1964) solution in reconnection theory.

\newpage

{\sl Acknowledgements}: We are grateful to I. V. Kubyshkin for
helpful discussions. Part of this work was done while VSS and DBK
were on a research visit to Graz. This work is partially supported
by the Russian Foundation of Basic Research , grant No.
\mbox{01-05-64954}, by the INTAS-ESA grant No. 99-01277, and by
the programme INTERGEOPHYSICS from the Russian Ministry of Higher
Education.  Part of this work is supported by the Austrian ``Fonds
zur F\"orderung der wissenschaftlichen Forschung'', project
P13804-TPH. We acknowledge support by the Austrian Academy of
Sciences, ``Verwaltungsstelle f\"ur Auslandsbeziehungen''.

\pagebreak

\newpage

\setcounter{figure}{0} 

\section*{Figure Captions}
\begin{enumerate}
\item Magnetic flux tube for the case $\omega_1=1,\omega_2=1$.
\item Surface $\beta=const$ for the case $\omega_1=1,\omega_2=1$.
\item Surface $\alpha=const$ for the case $\omega_1=1,\omega_2=1$.
\item Two linked flux tubes for the case $\omega_1=1,\omega_2=1$.
\item Surface $\psi=const$ for the case $\omega_1=1,\omega_2=1$.
\item Magnetic flux tube for the case $\omega_1=2,\omega_2=1$.
\item Surface $\alpha=const$ for the case $\omega_1=2,\omega_2=1$.
\item Two linked flux tubes for the case $\omega_1=2,\omega_2=1$.
\item Magnetic flux tube for the case $\omega_1=2,\omega_2=3$.
\item Surface $\alpha=const$ for the case $\omega_1=2,\omega_2=3$.
\item Central fragment of the surface $\alpha=const$ for the case $\omega_1=2, \omega_2=3$.
\item Two linked flux tubes for the case $\omega_1=2,\omega_2=3$.

\end{enumerate}


\begin{thebibliography}{17}



\bibitem{1}

  Biskamp,~D., {\it Nonlinear magnetohydrodynamics}, Cambridge University Press, 1993.



\bibitem{2}

  Chandrasekhar,~S., {\it Hydrodynamic and hydromagnetic stability}, Oxford University Press, 1961.





\bibitem{3}

  Crowell,~R. and R.~Fox, {\it Introduction to knot theory}, New York, 1963.



\bibitem{4}

  Dubrovin,~B.~A., S.~P.~Novikov, and A.~T.~Fomenko, {\it Modern geometry}, Nauka, Moscow, 1979.



\bibitem{5}

  Harris,~E.~G., On a plasma sheath separating regions of oppositely directed
  magnetic fields, {\it Nuovo Cimento}, {\bf 23}, 115, 1962.



\bibitem{6}

  Kamchatnov,~A.~M., Topological soliton in magnetohydrodynamics,
  {\it Sov. JETP}, {\bf 82}, No 1, 117, 1982.



\bibitem{7}

  Moffat,~H.~K. , {\it Magnetic field generation in electrically conducting fluids},
   Cambridge University Press, 1978.





\bibitem{8}

Petschek, H. E., Magnetic field annihilation, {\sl NASA Spec.
Publ.}, {\bf SP--50}, 425, 1964.



\bibitem{9}

 Pudovkin,~M.~I. and V.~S.~Semenov, Magnetic field reconnection theory and the
solar wind--magnetosphere interaction: A review, {Space Sci.
Revs.}, {\bf 41}, 1, (1985).



\bibitem{10}

  Priest,~E. and T.~Forbes, {\it Magnetic reconnection}, Cambridge University Press, 2000.



 \bibitem{11}

  Pustovitov,~V.~D., Magnetic coordinates with double straightening,
  {\it Plasma Phys. Rep.}, {\bf 25}, No 12, 963, 1999.



\bibitem{12}

  Sagdeev,~R.~Z., S.~S.~Moiseev, A.~V.~Tur, and V.~V.~Yanovskii, Problems of the theory
  of strong turbulence and topological soliton, in {\it Nonlinear Phenomena in Plasma
  Physics and Hydrodynamics}, edited by  R.~Z.~Sagdeev, (MIR Publishes, Moscow, 1986), p.135.



\bibitem{13}

  Schindler,~K., M.~Hesse, and J.~Birn, General magnetic reconnection, parallel electric
  fields and helicity, {\it J. Geophys. Res.}, {\bf 93}, No A6, 5547, 1988.





\bibitem{14}

  Taylor,~J.~B., Relaxation revisited, {\it Phys. Plasmas}, {\bf 7}, No 5, 1623, 2000.



\bibitem{15}

  Wiegelmann,~T. and J.~B\"uchner, Evolution of magnetic helicity in the course
  of kinetic magnetic reconnection, {\it Nonlinear Processes in Geophysics},
   {\bf 8}, No 3, 1623, 2001.



\bibitem{16}

  Wong,~H.~V., Particle canonical variables and guiding center Hamiltonian up
  to second order in the Larmor radius, {\it Phys. Plasmas}, {\bf 7}, No 1, 73, 2000.



\bibitem{17}

  Yoshizawa,~A., N.~Yokoi, and H.~Kato, Collimation mechanism of
  magnetohydrodynamic jets based on helicity and cross-helicity
  dynamos, with reference to astronomical jets,

  {\it Phys. Plasmas}, {\bf 7}, No 6, 2646, 2000.



\end{thebibliography}
\end{document}